\documentclass[11pt]{article}

\usepackage{arxiv}

\usepackage[utf8]{inputenc}
\usepackage[T1]{fontenc}
\usepackage{amsmath,amssymb,amsthm}
\usepackage{graphicx}
\usepackage{hyperref}
\usepackage{booktabs}
\usepackage{algorithm}
\usepackage{algorithmicx}
\usepackage{algpseudocode}
\usepackage{makecell}
\usepackage{xcolor}
\usepackage{natbib}
\usepackage{url}

\title{Fine-Tuning Autobidders with Group Relative Policy Optimization}

\author{
  Anton Safin \\
  Avito Research \\
  Moscow, Russia \\
  \And
  Alexandra Khirianova \\
  IAI MSU \& Avito Research \\
  Moscow, Russia \\
  \And
  Andrey Pudovikov \\
  IAI MSU \& Avito Research \\
  Moscow, Russia \\
  \And
  Aleksandr Katrutsa\\
  Avito Research \& AI Center MSU \\
  Moscow, Russia \\
  \texttt{amkatrutsa@gmail.com}
  \And
  Egor Samosvat \\
  Avito Research \\
  Moscow, Russia \\
}

\date{}

\begin{document}

\maketitle
\begin{abstract}
Automated bidding (autobidding) is a core component of modern online advertising systems.
Within this component, advertisers delegate sequential bid decisions to algorithms that must maximize campaign value while adhering to constraints such as a limited budget and a target cost-per-click (CPC).
One of the approaches to resolve the autobidding problem is to formulate it as a Markov decision process and use reinforcement learning (RL) to train a bid generation function.
The standard RL framework is actor-critic, which consists of an actor network that generates actions and a critic network that estimates the value of those actions.
In our setting, the action is typically a bid or related pacing multiplier, and the value is the expected return from the auction given the bid.
However, the alternating training of actor-critic RL models leads to instability and reduced robustness to noise. 
To address these issues, we adapt the Group Relative Policy Optimization (GRPO) framework to the autobidding setting.
This framework is a \emph{critic-free} policy-gradient method originally developed for large language model post-training, where the ground-truth target is unknown. 
The autobidding setting shares this property, since the optimal bid is unknown in advance. 
Moreover, GRPO in the LLM domain is used to fine-tune the pre-trained model, and we use the same technique to enhance the performance of the strong heuristic baseline.
We empirically compare Autobidding GRPO with actor-critic models, simple heuristics, and controller-based methods on the BAT, iPinYou, and AuctionNet benchmarks.
Extensive experiments show that Autobidding GRPO consistently outperforms baselines in clicks and is the best or second-best method in conversion volume.



\end{abstract}


\newcommand{\add}[1]{\textcolor{blue}{#1}}
\newcommand{\ask}[1]{\textcolor{red}{#1}}

\section{Introduction}
Online advertising platforms allocate impressions through real-time auctions, most commonly first-price (FP) and Vickrey--Clarke--Groves (VCG) mechanisms~\cite{wang2017display}.
At the scale of modern marketplaces, advertisers cannot set competitive bids manually for every opportunity. 
Instead, they rely on \emph{autobidders} that continuously adjust bids to maximize campaign objectives subject to budget and efficiency constraints such as target CPC or return on ad spend (ROAS)~\cite{lee2013real,aggarwal2019autobidding}.
The quality of these resulting bid adjustment strategies directly affects advertiser return on investment, marketplace liquidity, and platform revenue.

Autobidding is inherently a sequential process since decisions early in a campaign change remaining budget and reshape feasible actions later, while market competition and traffic intensity vary over time.
These properties motivate reinforcement learning (RL) formulations based on Markov decision processes (MDPs) or constrained MDPs (CMDPs).
These formulations introduce a bidder, which observes campaign and market state, selects a bid (or a pacing multiplier), and receives stochastic rewards and costs from auction outcomes~\cite{cai2017real,wu2018budget,wu2022sustainable}.
Despite substantial progress, deploying RL autobidders remains challenging~\cite{korenkevych2024offline,wu2022sustainable}.
Actor--critic methods such as PPO~\cite{schulman2017ppo} train not only a bidding policy, but also a separate value network that estimates expected return.
Under long campaigns and noisy auctions, this value estimate is hard to learn reliably, which can make policy updates unstable.

Group Relative Policy Optimization (GRPO)~\cite{shao2024deepseekmath} was recently introduced as a critic-free alternative for RL post-training of large language models.
Instead of training a critic model that evaluates the suggestions from the actor model for each prompt, the GRPO approach samples a \emph{group} of outputs and computes advantages from relative rewards within that group.
This approach avoids the alternating training of actor-critic RL models, which is known to be unstable and prone to noise.
We propose adapting the GRPO approach to train an RL-based autobidding model and name the proposed approach \textbf{AB-GRPO}.
Indeed, the autobidding problem shares the same instances as the LLM post-training problem.
First, given a fixed campaign context (budget, lifetime, CPC target, and local market statistics), one can sample multiple bidding trajectories and compare their realized values together with constraint satisfaction.
Based on these observations, one updates the policy model (autobidder) toward relatively better trajectories without training a separate critic model.
Second, the GRPO approach to LLM post-training assumes a pre-trained LLM that can be further improved. 
AB-GRPO uses the Adaptive linear model~\cite{khirianova2025bat} as the tuned base autobidder and adjusts its parameters to improve performance on the given task.
Our study focuses on maximizing the number of clicks.
Thus, AB-GRPO trains a single policy that samples multiple bid trajectories for the base autobidder. 
By design, AB-GRPO requires fewer computational resources and avoids the alternating training inherent to actor-critic models, leading to greater training stability.

The main contributions of our work are summarized below
\begin{enumerate}
  \item We propose the AB-GRPO method, which leverages the advantages of the GRPO approach for LLM post-training and adapts it to the constrained autobidding problem.
  \item We develop the AB-GRPO pipeline for training a single policy that samples multiple trajectories for the base autobidder without training a separate critic model.
  \item We evaluate the proposed AB-GRPO method on three benchmark datasets and show that it appears to be the best or the second-best method compared to simple heuristics, controller-based autobidders, and actor-critic RL methods.
\end{enumerate}

\section{Related works}




Reinforcement Learning has emerged as a natural paradigm for sequential decision-making in RTB environments. Early works applied policy gradient methods to learn bidding strategies directly from auction feedback~\cite{cai2017reinforcement}. However, applying standard on-policy RL methods like PPO~\cite{schulman2017ppo} to RTB presents significant challenges: PPO is sample-inefficient, suffers from sparse and noisy rewards that induce high gradient variance, struggles with non-stationary market dynamics, and requires a separate critic network that doubles memory and computational costs.


To address the computational burden of learned critics, Shao et al.~\cite{shao2024deepseekmath} introduced Group Relative Policy Optimization (GRPO) in the context of mathematical reasoning for Large Language Models. Instead of learning an explicit value function, GRPO generates a cohort of candidate actions from the same state and computes relative advantages via intra-group reward normalization, effectively eliminating the need for a separate critic network.


Recent work has systematically examined GRPO's properties. Lian et al.~\cite{lian2025comparative} presented a controlled comparison of PPO, GRPO, and DAPO for LLM reasoning enhancement, finding that increasing group size leads to more stable training dynamics and higher accuracy, while the impact of the KL-penalty coefficient is non-monotonic. Oliveira et al.~\cite{oliveira2025learning} conducted the first systematic evaluation of GRPO in classical single-task RL environments, revealing that learned critics remain essential for long-horizon tasks—GRPO underperforms PPO except in short-horizon settings—and that smaller group sizes can outperform larger ones in certain continuous control tasks. These findings inform our sensitivity analysis and highlight the importance of task-specific tuning.


The application of GRPO to autobidding has recently gained attention. Huang et al.~\cite{huang2025generative} introduced Generative Bid Shading (GBS), which comprises an end-to-end generative model for shading ratios and a reward preference alignment system that optimizes short-term and long-term surplus using GRPO. GBS has been deployed on the Meituan DSP platform, serving billions of bid requests daily. However, GBS addresses a fundamentally different problem—\emph{bid shading} at the per-request level—where the goal is to reduce the bid to avoid overspending on individual impressions while maintaining win rate.

The DARA framework~\cite{dara2026} introduced GRPO-Adaptive, an LLM post-training strategy for few-shot budget allocation across multiple advertising channels. DARA dynamically updates the reference policy during GRPO training to enhance reasoning and numerical precision.



Our work occupies a distinct position in the autobidding landscape. Table~\ref{tab:baselines_comparison} summarizes the key differences between our approach and the most closely related GRPO-based systems.

\begin{table}[!hbt]
\centering
\caption{Comparison of GRPO-based autobidding approaches.}
\begin{tabular}{lccc}
\toprule
 & \textbf{GBS}~\cite{huang2025generative} & \textbf{DARA}~\cite{dara2026} & \textbf{Our work} \\
\midrule
Approach & Bid shading & Budget allocation & Budget pacing  \\
Granularity & Per-impression & Per-campaign & Per-hour \\
GRPO role & \makecell{Full policy\\optimization} & \makecell{Adaptive\\reference update} & \makecell{Delta-correction\\to linear baseline} \\
\bottomrule
\end{tabular}
\label{tab:baselines_comparison}
\end{table}

To the best of our knowledge, our work is the first to apply GRPO as a corrective enhancement for budget pacing in second-price auctions, and to provide a direct head-to-head comparison between GRPO and actor-critic RL methods in this setting.

\section{Problem statement: autobidding as RL}
\label{sec:problem}

\subsection{Campaign-level bidding objective}

Consider an advertising campaign with lifetime horizon $T$ (hours),
total budget $B$, and an optional target average CPC $C$.
At each hour $t=1,\ldots,T$, the campaign participates in a set of auctions
$\mathcal{I}_t$. 
For auction $i\in\mathcal{I}_t$, the autobidder submits a bid $b_{t,i}\ge 0$.
Let
\begin{equation}
\mathrm{win}_{t,i}
=
\begin{cases}
1, & \text{if the campaign wins auction $i$ at hour $t$},\\
0, & \text{otherwise},
\end{cases}
\end{equation}
and let $wp_{t,i}\ge 0$ be the clearing/payment price induced by the auction rule.
The incurred cost is
\begin{equation}
\mathrm{Cost}_{t,i}
=
\mathrm{win}_{t,i}\, wp_{t,i}.
\end{equation}
If the auction is won, a click and a conversion may occur; we write $\mathrm{click}_{t,i}\in\{0,1\}$ and $\mathrm{cnv}_{t,i}\in\{0,1\}$, with $\mathrm{click}_{t,i}=\mathrm{cnv}_{t,i}=0$ whenever $\mathrm{win}_{t,i}=0$.
Following the constrained autobidding formulation used in recent benchmarks~\cite{pudovikov2025autobidding,khirianova2025bat,yang2019bid}, the campaign-level problem is
\begin{equation}
\label{eq:autobid-opt}
\begin{aligned}
\max_{\{b_{t,i}\}} \quad
& \mathbb{E}\!\left[
\sum_{t=1}^{T}\sum_{i\in\mathcal{I}_t} \mathrm{click}_{t,i}
\right] \\
\text{s.t.} \quad
& \sum_{t=1}^{T}\sum_{i\in\mathcal{I}_t}
\mathrm{win}_{t,i}\, wp_{t,i}
\le B, \\[0.25em]
& \frac{
\sum_{t=1}^{T}\sum_{i\in\mathcal{I}_t}
\mathrm{win}_{t,i}\, wp_{t,i}
}{
\sum_{t=1}^{T}\sum_{i\in\mathcal{I}_t}
\mathrm{click}_{t,i}
}
\le C
 \\[0.25em]
& b_{t,i}\ge 0,\qquad t=1,\ldots,T,\; i\in\mathcal{I}_t.
\end{aligned}
\end{equation}
The expectation is taken over auction competition, user responses, and traffic.
The first constraint is the budget limit; the second constraint is the average CPC constraint.
An analogous objective with $\mathrm{cnv}_{t,i}$ in place of $\mathrm{click}_{t,i}$ corresponds to conversion maximization.

\subsection{MDP formulation}

We model sequential bidding as an MDP
$\mathcal{M} = (\mathcal{S}, \mathcal{A}, P, r, \gamma)$
(or a CMDP when constraints are treated explicitly).
\begin{description}
  \item[State $s_t \in \mathcal{S}$.]
  A compact representation of campaign progress and market context, typically including
  remaining budget~$B_t$, remaining time $T-t$, cumulative value and spend, predicted
  CTR, and local traffic or competition signals.
  \item[Action $a_t \in \mathcal{A}$.]
  Either a bid $b_t$ or a pacing multiplier mapped to bids for opportunities
  within tick $t$. 
  We consider continuous action spaces, which are standard for bid control.
  \item[Transition probability $\mathbb{P}(s_{t+1}\mid s_t,a_t)$.]
  Induced by auction clearing, stochastic clicks/conversions, budget spending, and exogenous traffic dynamics.
  \item[Reward $r(s_t,a_t)$.]
  Instantaneous utility derived from acquired value, optionally shaped with penalties for
  constraint violations or pacing irregularity. A simple instantiation is
  $r_t = v_t - \eta\, c_t$, while constrained variants use Lagrangian rewards
  $r_t = v_t - \lambda_B c_t - \lambda_C \xi_t$, where $\xi_t$ measures CPC/ROAS violation and
  dual variables $(\lambda_B,\lambda_C)$ are adapted during training.
\end{description}
The learning objective is to find a policy $\pi_\theta(a_t\mid s_t)$ maximizing expected return
\begin{equation}
\label{eq:rl-obj}
J(\theta)
=
\mathbb{E}_{\tau \sim \pi_\theta}
\left[
\sum_{t=0}^{T-1} \gamma^t r(s_t,a_t)
\right],
\end{equation}
subject to (soft or hard) satisfaction of budget and CPC constraints over an episode.
An episode $\tau = (s_0,a_0,r_0,\ldots,s_T)$ corresponds to one campaign realization under a fixed context $q$ (campaign type, budget, lifetime, target CPC, and auction environment seed/statistics).

\subsection{Why GRPO is a natural fit}
\label{sec:why-grpo}

Actor--critic methods estimate advantages via a learned value function $V_\phi(s)$.
In autobidding, $V_\phi$ is difficult to fit reliably because returns depend on long-horizon
budget coupling and highly stochastic auction noise.
GRPO replaces the critic with a group baseline:
for a shared context $q$, sample a group of $G$ trajectories
$\{\tau_i\}_{i=1}^{G} \sim \pi_{\theta_{\mathrm{old}}}(\cdot\mid q)$,
assign each trajectory a scalar score $R_i$ (e.g., total value with constraint penalties),
and define the group-relative advantage
\begin{equation}
\label{eq:grpo-adv}
\hat{A}_i
=
\frac{R_i - \mathrm{mean}(\{R_j\}_{j=1}^{G})}
{\mathrm{std}(\{R_j\}_{j=1}^{G})+\varepsilon}.
\end{equation}
Policy updates then follow a clipped importance-weighted objective analogous to PPO, but
without a value network~\cite{shao2024deepseekmath}.
This matches the autobidding evaluation pattern of comparing alternative bidding behaviors
under identical campaign conditions, and reduces memory/compute overhead relative to
critic-based RL---a desirable property for large-scale offline simulation and repeated
campaign-level training.

\section{AB-GRPO framework}

In this section, we introduce the AB-GRPO framework which adapts to the autobidding domain the original GRPO framework, proposed for fine-tuning LLMs~\cite{shao2024deepseekmath}. 
We detail the modifications to the policy representation, state and action spaces, objective function, and training loop, aligning the algorithm with the campaign-level MDP defined in Section~\ref{sec:problem}.

\subsection{Adaptive linear model baseline}
Since the original GRPO framewotk is used for post-training of LLM, we consider the Adaptive Linear Model (ALM) as a baseline for autobidding that we can further improved via the GRPO mechanism. 
For the reader convenience and completeness, we briefly introduce the ALM technique here.

\textbf{ALM}~\cite{khirianova2025bat} is a traffic-aware pacing controller, which considers budget spends proportional to the traffic during the campaign lifetime.
Formally, ALM estimates bid through the log-scale prices
\[
  b_t=\gamma^{\mathrm{bin}_t},
\]
where $\gamma = 1.2$ in our experiments and  
\begin{equation}
  \mathrm{bin}_t
  =
  \mathrm{clip}\bigl(
  \mathrm{bin}_{t-1}+f\cdot\hat{B}^{\mathrm{end}}_t,\;
  \mathrm{bin}_{t-1}-\ell,\;
  \mathrm{bin}_{t-1}+u
  \bigr).
  \label{eq::alm_bin_t}
  \end{equation}
Here, the gain $f$ and the clip values $\ell,u$ are hyperparameters, which we tune for initial baseline preparing.
More interesting ingredient is $\hat{B}^{\mathrm{end}}_t$ since it is the forecast of remaining budget at the campaign end.
ALM uses the following formulas to estimate this quantity:
\begin{equation}
k_t
=
\frac{\hat{B}_t-\hat{B}_{t-1}}{tr(t_0,t)-tr(t_0,t{-}1)},
\qquad
\hat{B}^{\mathrm{end}}_t
=
\hat{B}_t+k_t\cdot tr(t,T).
\end{equation}
where $\hat{B}_t=B_t/B_0$ is the remaining-budget fraction,  $B_t$ is the available budget at time step $t$, $B_0$ is the initial campaign budget, $k_t$ the local spend slope and $tr(t_1, t_2)$ represents the fraction of total traffic expected between times $t_1$ and $t_2$.
A cold start $b_0=\kappa B_0$ is used when no history exists.

\textbf{AB-GRPO} uses fixed pre-tuned ALM parameters and trains policy that outputs a correction $\delta_t$ on the gain $f^*$ such that 
\[
f_t=f^*+\delta_t
\]
is used in~(\ref{eq::alm_bin_t}) for the $\mathrm{bin}_t$ estimation and directly affects bid.
The details of this adaptation are presented in the next section.

\subsection{Policy and State–Action Representation}

The GRPO policy for LLM post-training is implemented as an autoregressive Transformer that generates discrete token sequences. 
In contrast, we implement the GRPO policy for autobidding through a simple, lightweight MLP parametrized with trainable parameters~$\theta$.
This MLP model maps the input state $s_t$ at hour~$t$ to the mean $\mu_{\theta}(s_t)$ and variance $\sigma^2_{\theta}(s_t)$ of a normal distribution, which are then transformed into an action $a_t$.
The state $s_t \in \mathbb{R}^8$ at hour $t$ is constructed from normalised campaign features:
\begin{equation}
    \begin{split}
s_t = \Big[&
\hat{B}_t,
\frac{t-t_0}{T-t_0},
\frac{tr(t_0,t)}{tr(t_0,T)},
\frac{\hat B^s_t}{tr^s_t}, \text{bin}(b_{t-1}), \\
& \hat{B_t} + \hat B^s_t \cdot tr(t,T),
\text{CTR}_t,
\frac{\log(1+B_0)}{10}
\Big]^\top, \\
\text{where } & \hat B^s_t = \hat B_t -\hat  B_{t-1}, \quad tr^s_t = tr(t,t+1)- tr(t-1,t),
    \end{split}
\end{equation}
The variables $t_0$ and $T$ denote the campaign start and finish times, respectively. 
The step-level changes are defined by $B^s_t$ and $tr^s_t$ as shown above. 
Finally, $b_{t-1}$ is the bid submitted in the previous step, $\text{bin}(b)$ is its logarithmic bin index, and $\text{CTR}_t$ is the predicted click-through rate for the current period.

The action $a_t \equiv \delta_t$ is a bounded correction to the base bid factor produced by the pre-tuned linear ALM controller. 
To enforce the bound $\delta_t \in [-\Delta_{\max}, \Delta_{\max}]$ ($\Delta_{\max}=1.5$) while allowing Gaussian exploration, we use the squashed Gaussian parametrisation~\cite{haarnoja2018soft}:
\begin{equation}
z_t \sim \mathcal{N}\big(\mu_\theta(s_t), \sigma^2_\theta(s_t)\big),\qquad
\delta_t = \Delta_{\max} \cdot \tanh(z_t).
\end{equation}
To ensure gradient propagation through the $\pi_{\theta}$ model, we use the reparameterization trick to generate $z_t$. 
The bid $b_t$ is calculated as
\begin{equation}\label{eq:bid_calc}
\begin{split}
&f_t = f^* + \delta_t\\
&\mathrm{bin}_t
=
\mathrm{clip}(
\mathrm{bin}_{t-1}+f_t\cdot\hat{B}^{\mathrm{end}}_t,
\mathrm{bin}_{t-1}-\ell,
\mathrm{bin}_{t-1}+u) \\
&b_t = \gamma^{\mathrm{bin}_t},
\end{split}
\end{equation}
where $f$ is the pre-tuned and fixed ALM gain,
$\delta_t$ is the learned correction, and $\hat{B}^{\mathrm{end}}_t$ is the ALM forecast of the remaining-budget fraction at the campaign end.

The training of the GRPO policy $\pi_{\theta}$ requires the log-likelihood of a sampled action $\delta_t$, which can be computed via the change-of-variables formula:
\begin{equation}
\begin{split}
    \log \pi_\theta(\delta_t \mid s_t) = & \log \mathcal{N}(z_t \mid \mu_\theta(s_t), \sigma_\theta(s_t)) - \\ &\log\!\Big(\Delta_{\max} \cdot (1 - \tanh^2(z_t))\Big).
\end{split}
\end{equation}
Based on the introduced bid generation procedure, we describe the objective function for GRPO training in the following section.



\subsection{Objective Function}

For a group of $G$ independent rollouts $\{\tau_i\}_{i=1}^G$ sampled from the old policy $\pi_{\theta_{\text{old}}}$, let $R_i$ be the scalar reward of trajectory $i$ (defined below). The group-relative advantage is:
\begin{equation}\label{eq:advantage}
A_i = \frac{R_i - \mu_R}{\sigma_R + \varepsilon},\quad
\mu_R = \frac{1}{G}\sum_{g=1}^G R_g,\quad
\sigma_R = \sqrt{\frac{1}{G}\sum_{g=1}^G (R_g - \mu_R)^2}.
\end{equation}

For each transition $(s_t, a_t)$, we define the probability ratio
\begin{equation}
r_t(\theta) = \frac{\pi_\theta(a_t \mid s_t)}{\pi_{\theta_{\text{old}}}(a_t \mid s_t)}.
\end{equation}
The clipped surrogate loss is:
\begin{equation}
\mathcal{L}_{\text{policy}}(\theta) = - \frac{1}{|\mathcal{D}|} \sum_{(s_t,a_t)\in\mathcal{D}} \min\!\Big(r_t(\theta) A_i,\; \operatorname{clip}(r_t(\theta), 1-\epsilon, 1+\epsilon)\, A_i\Big),
\end{equation}
with $\epsilon = 0.2$.

Following the original GRPO formulation~\cite{shao2024deepseekmath}, we adopt a KL penalty against a reference policy $\pi_{\text{ref}}$ (initialised as a copy of $\pi_\theta$ before training and refreshed after each epoch). For diagonal Gaussians, the KL divergence is analytical:

\begin{equation}
D_{\mathrm{KL}}\big(\pi_\theta(\cdot\mid s) \,\|\, \pi_{\text{ref}}(\cdot\mid s)\big)
= \log\frac{\sigma_{\text{ref}}(s)}{\sigma_\theta(s)} + \frac{\sigma_\theta(s)^2 + (\mu_\theta(s)-\mu_{\text{ref}}(s))^2}{2\sigma_{\text{ref}}(s)^2} - \frac{1}{2}.
\end{equation}
The KL loss is averaged over the batch:
\begin{equation}
\mathcal{L}_{\mathrm{KL}}(\theta) = \frac{1}{|\mathcal{D}|} \sum_{(s_t,\cdot)\in\mathcal{D}} D_{\mathrm{KL}}\big(\pi_\theta(\cdot\mid s_t) \,\|\, \pi_{\text{ref}}(\cdot\mid s_t)\big).
\end{equation}

An entropy bonus encourages exploration:
\begin{equation}
H(\pi_\theta(\cdot\mid s)) = \frac{1}{2} + \frac{1}{2}\log(2\pi) + \log \sigma_\theta(s),
\end{equation}
with the entropy loss
\begin{equation}
\mathcal{L}_{\text{entropy}}(\theta) = - \frac{1}{|\mathcal{D}|} \sum_{(s_t,\cdot)\in\mathcal{D}} H(\pi_\theta(\cdot\mid s_t)).
\end{equation}

The overall objective is:
\begin{equation}
\mathcal{L}(\theta) = \mathcal{L}_{\text{policy}}(\theta) + \beta\,\mathcal{L}_{\mathrm{KL}}(\theta) - \alpha\,\mathcal{L}_{\text{entropy}}(\theta),
\end{equation}
with $\beta = 0.01$ and $\alpha = 0.001$.

\subsection{Training Procedure}

Training follows an on-policy, episodic scheme. One update step consists of three stages.

\textbf{Group collection.} For a given campaign context, we generate $G$ independent rollouts using the current policy $\pi_{\theta_{\text{old}}}$ without updating it. Each rollout runs for the full horizon (up to $T$ hourly steps). At each step, we sample $\delta_t$, compute the bid $b_t$ via (5.4), simulate the auction, and store $(s_t, a_t, \log \pi_{\theta_{\text{old}}}(a_t\mid s_t))$. At the end of rollout $i$, we compute the reward $R_i$. In our default setup we use the total clicks:
\begin{equation}
R_i = \text{clicks}_i
\end{equation}
where $\text{clicks}_i$ is the total clicks obtained. We also experiment with a penalty for overspending, $R_i = (\text{clicks}_i - \lambda\,\text{spend}_i/B_0)/\log(1+B_0)$, and with a baseline-relative reward $\operatorname{clip}(\log(\text{clicks}_i+\alpha) - \log(\text{baseline\_clicks}+\alpha), -C, C)$. These rewards are used solely to compute the advantages $A_i$ via (5.7).

\textbf{Advantage computation and policy update.} After collecting all $G$ rewards, we compute $\mu_R$ and $\sigma_R$ as in (5.7) and assign to every transition in rollout $i$ the same advantage $A_i$. We then assemble the batch $\mathcal{D}$ with all transitions, compute the loss $\mathcal{L}(\theta)$ (5.14), and perform one or several (``PPO epochs'') gradient steps with the Adam optimiser. The collected data are then discarded.

We experiment with advantage reweighting, scaling each campaign’s within-group advantage by 
\begin{equation}\label{eq:advantage_scaling}
\hat{A}_i = \frac{A_i}{\log(B_0^{(i)})}
\end{equation}
after group standardization.

\textbf{Reference policy.} The reference policy $\pi_{\text{ref}}$ is initialised as a copy of $\pi_\theta$ before training and is refreshed after each full epoch.

The key hyperparameters are summarised in Table~\ref{tab:grpo-hyper}. This procedure eliminates the need for a learned critic, replacing it with a group-based empirical baseline—a desirable property in sparse, long-horizon settings like budget pacing.

\begin{table}[h]
\centering
\caption{GRPO hyperparameters used in our experiments.}
\begin{tabular}{lcc}
\toprule
Parameter & Symbol & Default \\
\midrule
Group size & $G$ & 16 \\
Clipping range & $\epsilon$ & 0.2 \\
KL coefficient & $\beta$ & 0.01 \\
Entropy coefficient & $\alpha$ & 0.001 \\
Max correction & $\Delta_{\max}$ & 1.5 \\
Learning rate & & $3\times10^{-4}$ \\
PPO update epochs & & 1 \\
Training horizon (hours) & $T$ & 24 \\
\bottomrule
\end{tabular}
\label{tab:grpo-hyper}
\end{table}

The overall training pipeline of AB-GRPO algorithm is also summarised in Algorithm~\ref{alg:grpo-alm} for completeness.

\begin{algorithm}
  \caption{AB-GRPO algorithm}
  \label{alg:grpo-alm}
  \begin{algorithmic}[1]
    \State \textbf{Initialize:} Policy $\pi_\theta$, reference $\pi_{\text{ref}} \leftarrow \pi_\theta$ \\ empty trajectory buffer $M$ and empty reward list $\mathcal{R}$
    \State \textbf{Given:} Group size $G$, clip $\epsilon$, KL coef $\beta$, entropy coef $\alpha$, horizon $H$, delta limit $\Delta_{\max}$, base factor $f$, small constant $\varepsilon$

    \For{epoch $e = 1$ to $E$}
      \For{each campaign $c$}
        \State $M \leftarrow \varnothing$, $\mathcal{R} \leftarrow \varnothing$
        \State Sample start hour $h \sim \mathcal{U}(0,\, T-H)$

        \For{$g = 1$ to $G$}
          \State Reset environment, $R_g \leftarrow 0$, $\tau_g \leftarrow \varnothing$
          \While{not terminal}
            \State Observe state $s_t$
            \If{$h \le t < h+H$}
              \State Sample action $a_t \sim \pi_\theta(\cdot|s_t)$
              \State Store $(s_t, a_t, \log \pi_\theta(a_t|s_t))$ in $\tau_g$
            \Else
              \State $a_t = 0$
            \EndIf
            \State Compute bid $b_t$ from $(s_t, a_t)$ using (\ref{eq:bid_calc})
            \State Execute $b_t$, observe reward $r_t$, $R_g \leftarrow R_g + r_t$
          \EndWhile
          \State $M \leftarrow M \cup \{\tau_g\}$, $\mathcal{R} \leftarrow \mathcal{R} \cup \{R_g\}$
        \EndFor
        \State Compute advantages $A_g$ according to (\ref{eq:advantage})

        \For{$p = 1$ to $P$}
          \State Compute $\log \pi_\theta$, entropy $\mathcal{H}$, and $KL = \mathcal{D}_{KL}(\pi_\theta \,\|\, \pi_{\text{ref}})$ over all transitions in $M$
          \State $\mathcal{L}_{\text{policy}} = -\mathbb{E}_{M}\!\left[\min\left(r(\theta)\, A_g,\; \operatorname{clip}(r(\theta), 1-\epsilon, 1+\epsilon)\, A_g\right)\right]$
          \State $\mathcal{L} = \mathcal{L}_{\text{policy}} + \beta\,\mathbb{E}_{M}[KL] - \alpha\,\mathbb{E}_{M}[\mathcal{H}]$
          \State Update $\theta$ by gradient optimizer on $\mathcal{L}$
        \EndFor
        \State $\pi_{\text{ref}} \leftarrow \pi_\theta$
      \EndFor
    \EndFor
  \end{algorithmic}
\end{algorithm}

\subsection{Algorithm features}

Many RL approaches to autobidding adopt the actor--critic paradigm, where a critic network estimates state values to reduce gradient variance. However, this introduces serious practical issues in autobidding: PPO~\cite{schulman2017ppo} \textbf{doubles memory and inference cost}, while the critic becomes \textbf{unstable under sparse, noisy, terminal-only rewards}. GRPO eliminates the critic entirely, computing advantages via group normalization of observed returns, which removes bias and stabilises training. Moreover, GRPO compares independent rollouts from the same context, directly mirroring autobidding's natural evaluation logic, and recent studies~\cite{lian2025comparative} confirm that \textbf{larger groups improve stability}. 

Rather than learning a policy from scratch, we train only a corrective delta $\delta_t$ applied to a \textbf{stable linear baseline}, preserving interpretability and budget safety. The action is clipped, guaranteeing that the correction never exceeds safe limits. 

\section{Numerical experiments}

This section presents the evaluation of the AB-GRPO algorithm performance. 
We also compare it with critic-based RL, or the strongest competitors in this aspect~\cite{pudovikov2025autobidding}. 
We repeat the comparison on public datasets: iPinYou \cite{ipinyou}, BAT \cite{khirianova2025bat} and AuctionNet \cite{su2024a} to test robustness across markets and auction formats.

All methods are tuned for the same objective. The primary metric is total clicks on an untouched test set, averaged
over 3 random seeds. 
Conversions, pacing error, and realized cost per click are secondary
trade-offs and are not tuned.

\subsection{Datasets}
\label{sec:datasets}
To evaluate the performance of our AB-GRPO algorithm we consider three benchmark datasets: iPinYou, BAT, and AuctionNet.
Below we provide a brief description of each dataset and features important for experiment design.

\textbf{iPinYou}~\cite{ipinyou} is a single-slot second-price log with 5 advertisers from Season~2 and $N=35$ campaigns.
A campaign is one advertiser on one calendar day ($T=24$).
The train/val/test split is $20/10/5$ ($\approx 50\%/36\%/14\%$).
The auction budget $B_0$ is half of the logged daily spend.
Bids are updated every hour. The log has no predicted CTR/CVR, so both are Laplace-smoothed rates from that hour's realized clicks, impressions, and conversions.

Unlike impression-level auction logs such as iPinYou,  \textbf{BAT}~\cite{khirianova2025bat} contains aggregated auction statistics from a classified marketplace, with $N=1882$ campaigns after the 24-hour filter. 
The train/val/test split is $743/209/930$ ($\approx 39\%/11\%/49\%$).
Bids are updated every hour. 
Predicted CTR and conversion rates are provided in the log.

\textbf{AuctionNet}~\cite{su2024a} is a multi-slot second-price log with $S=48$ sellers.
We use two equal periods of $T=48$ steps each: the first for tuning and the second for test similarly to~\cite{pudovikov2025autobidding}.
The budget is $B_0=10^{4}$ per seller.
Bids are updated every 30 minutes. 

\subsection{Metrics}

Hyperparameters are chosen by total clicks on validation are then reported on the test set.

Let campaigns be indexed by $s$, hours by $t=1,\ldots,T$, and auctions
within hour $t$ by $i\in\mathcal{I}_{s,t}$.
Using the notation of Section~\ref{sec:problem}, define the aggregated
hourly outcomes
\begin{equation}
\begin{aligned}
c_{st}
&=
\sum_{i\in\mathcal{I}_{s,t}} \mathrm{click}_{s,t,i},
\qquad
v_{st}
=
\sum_{i\in\mathcal{I}_{s,t}} \mathrm{cnv}_{s,t,i},
\\
\mathrm{Cost}_{st}
&=
\sum_{i\in\mathcal{I}_{s,t}}
\mathrm{win}_{s,t,i}\, wp_{s,t,i}.
\end{aligned}
\end{equation}
Under first-price, $wp_{s,t,i}$ equals the submitted bid on a win;
under second-price/VCG, it is the logged clearing price.
The volume metrics are
\begin{equation}
\mathrm{Clicks}
=
\sum_{s}\sum_{t=1}^{T} c_{st},
\qquad
\mathrm{Cnv}
=
\sum_{s}\sum_{t=1}^{T} v_{st}.
\end{equation}
Clicks is the tuning objective.
Conversions are a secondary trade-off and are not tuned.
The realized cost per click is
\begin{equation}
\mathrm{CPC}
=
\frac{\sum_{s}\sum_{t=1}^{T}\mathrm{Cost}_{st}}
{\sum_{s}\sum_{t=1}^{T} c_{st}}.
\end{equation}

Budget pacing metric is the RMSE between realized spend and a target spend, averaged over campaigns:
\begin{equation}
\mathrm{RMSE}
=
\frac{1}{N}
\sum_{i}
\sqrt{\frac{1}{T}\sum_{t=1}^{T}\bigl(B_{*}^{(t)}-B_{i}^{(t)}\bigr)^{2}},
\qquad
B_{*}^{(t)}
=
\frac{B_0\,(T-t)}{T}.
\end{equation}
On AuctionNet the target remaining budget is taken linear in time. On BAT and iPinYou datasets the target spend in hour $t$ is $B_0$ times that hour's traffic share, and RMSE is divided by the mean hourly budget~\cite{khirianova2025bat}.

\subsection{Baselines}

The comparison uses the strongest published autobidders from each standard RTB family: a linear rule, a traffic-pacing controllers, and actor--critic RL. Critic-free RL algorithm GRPO, described in Section~4, is not a separate bidding rule: it searches a residual on locked ALM.

\textbf{Linear}~\cite{zhang2014optimal}
is a static multiple of predicted click-through rate,
\begin{equation}
b_t^{\mathrm{lin}}
=
\alpha\cdot \mathrm{CTR}_t.
\end{equation}
The single hyperparameter $\alpha$ is held constant over the campaign.

\textbf{PID} and \textbf{M-PID}~\cite{yang2019bid}
keep the dual bid of Yang et al.,
\begin{equation}
b_t^{\mathrm{OPT}}
=
\frac{\mathrm{CTR}_t\cdot\mathrm{CVR}_t+\mathrm{CTR}_t\cdot C\cdot q_t}{p_t+q_t},
\end{equation}
where $(p_t,q_t)$ are dual variables for the budget and cost-per-click constraints.
PID updates $(p_t,q_t)$ from the pacing error and the realized cost per click.
M-PID mixes the two PID channels with a $2\times 2$ matrix before that update.

\textbf{USCB}~\cite{USCB}
starts from a CTR-linear bid
\begin{equation}
b_t^{\mathrm{USCB}}
=
w_t\cdot \mathrm{CTR}_t.
\end{equation}
Unlike Linear, the multiplier $w_t$ is not a fixed $\alpha$.
An actor--critic updates it by a residual $a_t$,
\begin{equation}
w_t = w_{t-1}(1+a_t).
\end{equation}

\textbf{FAB}~\cite{liu2020dynamicFAB}
also starts from a CTR-linear bid.
An actor--critic residual $a_t$ scales that base,
\begin{equation}
b_t^{\mathrm{FAB}}
=
\frac{\mathrm{CTR}_t\cdot b_0}{\overline{\mathrm{CTR}}\,(1+a_t)}.
\end{equation}
Here $b_0$ is a fixed base bid, not the campaign budget $B_0$, and $\overline{\mathrm{CTR}}$ is the mean predicted click-through rate over positive entries of the training log.

\subsection{Performance of the AB-GRPO algorithm}

GRPO fine-tunes a tuned controller: the ALM gain $f$, the clip widths, and the cold-start coefficient remain fixed, and only a bounded residual $\delta_t$ is trained, with $f_t=f+\delta_t$.
The same residual applies to any heuristic or controller that maps a small set of scalars to a bid replayable in the simulator.
On the Linear algorithm, the residual adjusts $\alpha$; on PID, it adjusts the duals $(p_t,q_t)$ or the PID gains.
That transfer is not evaluated here and is left to future work.
ALM is the base controller in this paper because it is a competitive pacing rule and is low-cost to learn to calibrate in isolation: four scalars, no critic, and a closed-form bid.

\paragraph{BAT}
The first benchmark we use for evaluation of our AB-GRPO against selected baselines is BAT dataset~\cite{khirianova2025bat}, see summary of the dataset statistics in Section~\ref{sec:datasets}.
Since the BAT environment is deterministic, we do not use multiple seeds for evaluation.

Table~\ref{tab:bat} reports the performance of the compared baselines and proposed AB-GRPO on the BAT dataset.
We observe that AB-GRPO outperforms comsidered competitors in terms the number of clicks and the volume of conversions. 
At the same time, the RMSE value of AB-GRPO is moderate although is not the smallest. 
The PID provides the smallest RMSE while the number of clicks and conversions are not the best or second-best.
Such mismatch between the number of clicks and the RMSE value confirms the previous observation in~\cite{pudovikov2025autobidding} that the RMSE metric is not the best indicator of the quality of the autobidding algorithm.

\begin{table}[!ht]
    \centering
    \caption{Performance of the considered autobidding algorithms on the BAT dataset. 
    Our AB-GRPO algorithm achieves the highest number of clicks and conversions compared to the baselines. RMSE and CPC are moderate and smaller than the base ALM method.}
    \begin{tabular}{lcccc}
    \toprule
         & \# clicks & \# conversions & RMSE & CPC \\
         \midrule
         Linear & 5577 & 637 & 1.302 & \textbf{13.45} \\
         ALM & \underline{7669} & \underline{757} & 1.709 & 51.22 \\
         \textbf{AB-GRPO} & \textbf{7764} & \textbf{766} & 1.707 & 43.90 \\
         M-PID & 6446 & 682 & 1.29 & 14.57 \\
         PID & 7179 & 732 & \textbf{1.179} & 22.88 \\
         USCB & 4880 & 497 & 1.89 & 48.36 \\
         FAB & 5629 & 639 & \underline{1.27} & \underline{13.88} \\
         \bottomrule
    \end{tabular}
\label{tab:bat}
\end{table}

\paragraph{iPinYou}
The next benchmark we use is iPinYou dataset~\cite{ipinyou}, see summary of the dataset statistics in Section~\ref{sec:datasets}.
Table~\ref{tab:ipinyou1} reports total number of clicks, conversions, RMSE and CPC values for campaigns on the test set averaged over three random seeds.
AB-GRPO provides the highest number of clicks and outperforms the base ALM bidder, the staic Linear bidder, controller-based bidders (PID, M-PID) and actor-critic RL methods (USCB, FAB).
Since the target metric in this experiment is the number of clicks for all baselines, the presented performance of the AB-GRPO highlights its advantage over the baselines in the target scenario.

At the same time, the AB-GRPO provides the second-highest number of conversions and matches the base ALM bidder in this metric.
Note that, the highest number of conversion is given by Linear bidder and FAB, while FAB gives much smaller number of clicks and therefore cost for them less (see CPC column).

Regarding the RMSE metric, the AB-GRPO provides moderate values smaller than ALM and USCB bidders, and larger than Linear and controller-based bidders.
Since controller-based bidders directly affects the budget pacing through the cumulative signal of spends, their RMSE performance is not surprising.
Overall, AB-GRPO improves the primary objective (clicks) over its ALM initialization without harming conversions and with a modest pacing gain.

\begin{table}[!ht]
    \centering
    \caption{Performance comparison of the compared baselines and proposed AB-GRPO on the iPinYou dataset.
    AB-GRPO provides the largest number of clicks and the second-largest volume of conversions compared to baselines. 
    The RMSE is moderate and smaller than the base ALM method.
    }
    \begin{tabular}{lcccc}
    \toprule
         & \# clicks & \# conversions & RMSE & CPC \\
         \midrule
         Linear & $932.2_{\pm 12.9}$ & $\mathbf{36.0}_{\pm 0.0}$ & $\mathbf{0.993}_{\pm 0.009}$ & $21655_{\pm 874}$\\
         ALM & $\underline{932.5}_{\pm 15.1}$ & $\underline{34.3}_{\pm 0.6}$ & $1.200_{\pm 0.035}$ & $26697_{\pm 4134}$\\
         \textbf{AB-GRPO} & $\mathbf{937.8}_{\pm 11.8}$ & $\underline{34.3}_{\pm 0.6}$ & $1.179_{\pm 0.020}$ & $26404_{\pm 3978}$\\
         M-PID & $836.2_{\pm 41.9}$ & $28.7_{\pm 5.1}$ & $1.112_{\pm 0.069}$ & $16043_{\pm 1272}$\\
         PID & $858.6_{\pm 13.3}$ & $32.3_{\pm 1.2}$ & $1.058_{\pm 0.021}$ & $15932_{\pm 1292}$\\
         USCB & $638.0_{\pm 117.4}$ & $28.7_{\pm 6.7}$ & $1.282_{\pm 0.074}$ & $\mathbf{2706}_{\pm 1684}$\\
         FAB & $813.7_{\pm 15.5}$ & $\mathbf{36.0}_{\pm 0.0}$ & $\underline{1.057}_{\pm 0.014}$ & $\underline{13344}_{\pm 1868}$\\
         \bottomrule
    \end{tabular}
\label{tab:ipinyou1}
\end{table}

\paragraph{AuctionNet}
The last but not least benchmark we use is AuctionNet dataset, see summary of the dataset statistics in Section~\ref{sec:datasets}.
Table~\ref{tab:auction-net} reports the performance of the considered baselines and proposed AB-GRPO on the AuctionNet dataset.
AB-GRPO outperforms the base ALM method and actor-critic RL methods (USCB, FAB) in terms of number of clicks and conversions.
This observation is aligned with the previous results on the performance of AB-GRPO on the BAT and iPinYou datasets.
RMSE and CPC metrics are also in the middle of the range of the considered baselines, and the position of AB-GRPO is consistent with the previous experiments.


\begin{table}[!ht]
    \centering
    \caption{Performance comparison on the AuctionNet dataset}
    \begin{tabular}{lcccc}
        \toprule
        & \# clicks & \# conversions & RMSE & CPC \\
        \midrule
        Linear & 19\,228.27 & \textbf{870.32} & 1\,393.13 & 0.5104 \\
        ALM    & 14\,194.35 & 629.06 & 1\,823.64 & 0.7175 \\
        \textbf{AB-GRPO}   & \underline{19\,677.79} & \underline{869.08} & 1\,297.49 & 0.4997 \\
        M-PID  & 19\,119.67 & 772.19 & \textbf{802.01} & 0.5162 \\
        PID    & \textbf{19\,784.77} & 839.10 & \underline{1\,083.68} & 0.4987 \\
        USCB   & 13\,536.19 & 608.28 & 1\,249.18 & \textbf{0.4727} \\
        FAB    & 18\,781.06 & 808.45 & 1\,090.29 & \underline{0.4756} \\
        \bottomrule
    \end{tabular}
    \label{tab:auction-net}
\end{table}

\subsection{Complexity comparison}
This section presents the complexity of training/tuning and inference of the considered autobidding algorithms.
We measure the trainin/tuning complexity as the runtime required to train neural networks in RL-based methods or tuning hyperparameters in controller-based and heuristic methods.
This feature of autobidding algorithms affects how often an algorithm can be retrained/retuned if the market environment changes.
Similarly, we measure inference complexity as the runtime required to process a predefined number of campaigns.
This feature indicates whether the autobidding algorithm is feasible for deployment given production latency limits.

Table~\ref{tab:complexity} reports the complexity of the considered autobidding algorithms.
We observe that AB-GRPO is much faster at inference than actor-critic RL alternatives and requires fewer computational resources to train the policy model.
It is still slower than the ALM base model and the Linear heuristic, which is natural due to the absence of the additional policy model in these baselines.


\begin{table}[!ht]
    \centering
    \caption{Complexity comparison of the considered models. Aligned CPU runtime comparison. 
    Inference is the sum of equivalent time for available period-9 bidding calls from the AuctionNet dataset; training/tuning is one period-7/8 trial. }
    \begin{tabular}{lcc}
        \toprule
        Algorithm & Inference runtime (s) & Training/tuning per trial (s) \\
        \midrule
        Linear & 0.08 & 28.67  \\
        ALM & 0.10 & 28.97  \\
        \textbf{AB-GRPO} & 0.44 & 315.30 \\
        USCB & 1.89 & 1054.18 \\
        FAB & 1.90 & 1247.36 \\
        \bottomrule
    \end{tabular}
    \label{tab:complexity}
\end{table}

\section{Conclusion and future work}
We have considered the autobidding problem for second-price auctions and proposed the GRPO-based RL algorithm to solve it.
Our algorithm is a critic-free alternative to existing actor-critic RL autobidding algorithms, which makes it more stable and efficient.
The key idea is similar to LLM post-training and includes taking the ALM controller as a pre-tuned model for further correcting its factors based on the trained policy.
This strategy is straightforward and easy for policy training since it does not include a separate critic model by design.
We have compared our AB-GRPO algorithm with the Linear heuristic, controllers, and actor-critic RL algorithms on three public datasets (BAT, iPinYou, and AuctionNet).
The results consistently confirm that AB-GRPO is the best or second-best algorithm in terms of clicks and conversions.
It consistently outperforms the actor-critic RL baselines (USCB, FAB), supporting the claim that the GRPO approach is a promising direction for the autobidding problem.
Moreover, the complexity comparison shows that the AB-GRPO algorithm is faster at inference than actor-critic RL methods and is comparable at training/tuning to controller-based bidders.

This work takes the first step in transferring the GRPO framework through budget pacing to the autobidding domain.
We have taken the ALM controller as the base model and trained a policy to correct its gain factor.
Extending this approach to other controllers and heuristics is straightforward and a natural direction for future work.
Other promising directions include extending the AB-GRPO algorithm to other auction settings and online evaluation in real-world auction environments.

\bibliographystyle{unsrt}
\bibliography{lib}

@inproceedings{wang2017display,
  title={{Display Advertising with Real-Time Bidding ({RTB}) and Behavioural Targeting}},
  author={Wang, Jun and Zhang, Weinan and Yuan, Shuai},
  booktitle={Foundations and Trends in Information Retrieval},
  year={2017}
}

@inproceedings{lee2013real,
  title={{Real Time Bid Optimization with Smooth Budget Delivery in Online Advertising}},
  author={Lee, Kuang-Chih and Jalali, Ali and Dasdan, Ali},
  booktitle={Proceedings of the Seventh International Workshop on Data Mining for Online Advertising},
  year={2013}
}

@inproceedings{zhang2014optimal,
  title={{Optimal real-time bidding for display advertising}},
  author={Zhang, Weinan and Yuan, Shuai and Wang, Jun},
  booktitle={Proceedings of the 20th ACM SIGKDD International Conference on Knowledge Discovery and Data Mining},
  pages={1077--1086},
  year={2014}
}

@inproceedings{yang2019bid,
  title={{Bid optimization by multivariable control in display advertising}},
  author={Yang, Xun and Li, Yasong and Wang, Hao and Wu, Di and Tan, Qing and Xu, Jian and Gai, Kun},
  booktitle={Proceedings of the 25th ACM SIGKDD International Conference on Knowledge Discovery \& Data Mining},
  pages={1966--1974},
  year={2019}
}

@inproceedings{USCB,
  author={He, Yue and Chen, Xiujun and Wu, Di and Pan, Junwei and Tan, Qing and Yu, Chuan and Xu, Jian and Zhu, Xiaoqiang},
  title={{A Unified Solution to Constrained Bidding in Online Display Advertising}},
  booktitle={Proceedings of the 27th ACM SIGKDD Conference on Knowledge Discovery \& Data Mining},
  pages={2993--3001},
  year={2021}
}

@article{liu2020dynamicFAB,
  title={{A dynamic bidding strategy based on model-free reinforcement learning in display advertising}},
  author={Liu, Mengjuan and Jiaxing, Li and Hu, Zhengning and Liu, Jinyu and Nie, Xuyun},
  journal={IEEE Access},
  volume={8},
  pages={213587--213601},
  year={2020}
}

@inproceedings{
su2024a,
title={{AuctionNet: A Novel Benchmark for Decision-Making in Large-Scale Games}},
author={Kefan Su and Yusen Huo and Zhilin Zhang and Shuai Dou and Chuan Yu and Jian Xu and Zongqing Lu and Bo Zheng},
booktitle={The Thirty-eight Conference on Neural Information Processing Systems Datasets and Benchmarks Track},
year={2024},
url={https://arxiv.org/abs/2412.10798}
}

@inproceedings{ipinyou,
  title={{iPinYou global rtb bidding algorithm competition dataset}},
  author={Liao, Hairen and Peng, Lingxiao and Liu, Zhenchuan and Shen, Xuehua},
  booktitle={Proceedings of the Eighth International Workshop on Data Mining for Online Advertising},
  pages={1--6},
  year={2014}
}

@inproceedings{aggarwal2019autobidding,
  title={{Autobidding with Constraints}},
  author={Aggarwal, Gagan and Badanidiyuru, Ashwinkumar and Mehta, Aranyak},
  booktitle={Web and Internet Economics (WINE)},
  pages={17--30},
  year={2019}
}

@inproceedings{cai2017real,
  title={{Real-Time Bidding by Reinforcement Learning in Display Advertising}},
  author={Cai, Han and Ren, Kan and Zhang, Weinan and Malialis, Kleanthis and Wang, Jun and Yu, Yong and Guo, Defeng},
  booktitle={Proceedings of the Tenth ACM International Conference on Web Search and Data Mining},
  year={2017}
}

@inproceedings{cai2017reinforcement,
  author    = {Cai, Han and Kan, Kan and Zhang, Weinan and Wang, Yong Yu and Yu, Hong and Wang, Jian},
  title     = {{Real-Time Bidding by Reinforcement Learning in Display Advertising}},
  booktitle = {Proceedings of the Tenth ACM International Conference on Web Search and Data Mining (WSDM)},
  pages     = {661--670},
  year      = {2017},
  publisher = {ACM},
  doi       = {10.1145/3018661.3018702}
}

@inproceedings{wu2018budget,
  title={{Budget Constrained Bidding by Model-free Reinforcement Learning in Display Advertising}},
  author={Wu, Di and Chen, Xiujun and Yang, Xun and Wang, Hao and Tan, Qing and Zhang, Xiaoxun and Xu, Jian and Gai, Kun},
  booktitle={Proceedings of the 27th ACM International Conference on Information and Knowledge Management},
  year={2018}
}

@inproceedings{wu2022sustainable,
  title={{Sustainable Online Reinforcement Learning for Auto-bidding}},
  author={Wu, Zhiyu and others},
  booktitle={Advances in Neural Information Processing Systems},
  year={2022}
}

@article{schulman2017ppo,
  title={{Proximal Policy Optimization Algorithms}},
  author={Schulman, John and Wolski, Filip and Dhariwal, Prafulla and Radford, Alec and Klimov, Oleg},
  journal={arXiv preprint arXiv:1707.06347},
  year={2017}
}

@article{shao2024deepseekmath,
  title={{{DeepSeekMath}: Pushing the Limits of Mathematical Reasoning in Open Language Models}},
  author={Shao, Zhihong and Wang, Peiyi and Zhu, Qihao and Xu, Runxin and Song, Junxiao and Bi, Xiao and Zhang, Haowei and Zhang, Mingchuan and Li, Y. K. and Wu, Y. and Guo, Daya},
  journal={arXiv preprint arXiv:2402.03300},
  year={2024}
}

@article{khirianova2025bat,
  title={{BAT: Benchmark for Auto-bidding Task}},
  author={Khirianova, Alexandra and Solodneva, Ekaterina and Pudovikov, Andrey and Osokin, Sergey and Samosvat, Egor and Dorn, Yuriy and Ledovsky, Alexander and Zenkova, Yana},
  journal={Proceedings of the ACM Web Conference 2025 (WWW '25)},
  year={2025}
}

@article{pudovikov2025autobidding,
  title={{Autobidding Arena: Unified Evaluation of Classical and RL-Based Autobidding Algorithms}},
  author={Pudovikov, Andrey and Khirianova, Alexandra and Solodneva, Ekaterina and Katrutsa, Aleksandr and Samosvat, Egor and Dorn, Yuriy},
  journal={arXiv preprint arXiv:2510.19357},
  year={2025}
}

@article{lian2025comparative,
  title={{Comparative Analysis and Parametric Tuning of PPO, GRPO, and DAPO for LLM Reasoning Enhancement}},
  author={Lian, Yongsheng},
  journal={arXiv preprint arXiv:2512.07611},
  year={2025}
}

@article{oliveira2025learning,
  title={{Learning Without Critics? Revisiting GRPO in Classical Reinforcement Learning Environments}},
  author={Oliveira, Bryan L. M. de and Frujeri, Felipe V. and Queiroz, Marcos P. C. M. and Martins, Luana G. B. and Soares, Telma W. de L. and Melo, Luckeciano C.},
  journal={arXiv preprint arXiv:2511.03527},
  year={2025}
}

@article{huang2025generative,
  title={{Generative Bid Shading in Real-Time Bidding Advertising}},
  author={Huang, Yinqiu and Ma, Hao and Chen, Wenshuai and Wang, Shuli and Zhang, Yongqiang and Wei, Xue and Zhu, Yinhua and Wang, Haitao and Wang, Xingxing},
  journal={arXiv preprint arXiv:2508.06550},
  year={2025}
}

@inproceedings{dara2026,
  title={{DARA: Few-shot Budget Allocation in Online Advertising via In-Context Decision Making with RL-Finetuned LLMs}},
  author={Song, Mingxuan and Huo, Yusen and Zhou, Bohan and Yin, Shenglin and Xiao, Zhen and Long, Jieyi and Zhang, Zhilin and Yu, Chuan},
  booktitle={Proceedings of the ACM Web Conference 2026 (WWW '26)},
  year={2026},
  note={arXiv:2601.14711}
}

@InProceedings{haarnoja2018soft,
  title = 	 {{Soft Actor-Critic: Off-Policy Maximum Entropy Deep Reinforcement Learning with a Stochastic Actor}},
  author =       {Haarnoja, Tuomas and Zhou, Aurick and Abbeel, Pieter and Levine, Sergey},
  booktitle = 	 {Proceedings of the 35th International Conference on Machine Learning},
  pages = 	 {1861--1870},
  year = 	 {2018},
  editor = 	 {Dy, Jennifer and Krause, Andreas},
  volume = 	 {80},
  series = 	 {Proceedings of Machine Learning Research},
  publisher =    {PMLR},
}

@inproceedings{korenkevych2024offline,
author = {Korenkevych, Dmytro and Cheng, Frank and Balakir, Artsiom and Nikulkov, Alex and Gao, Lingnan and Cen, Zhihao and Xu, Zuobing and Zhu, Zheqing},
title = {{Offline Reinforcement Learning for Optimizing Production Bidding Policies}},
year = {2024},
publisher = {Association for Computing Machinery},
booktitle = {Proceedings of the 30th ACM SIGKDD Conference on Knowledge Discovery and Data Mining},
pages = {5251–5259},
numpages = {9},
}
\end{document}